\documentclass[10pt]{article}
\usepackage{amssymb,dsfont,amsmath,amsfonts,bm,mathabx}
\usepackage[active]{srcltx}
\usepackage{authblk}
\usepackage{slashed}
\usepackage{color}
\usepackage{graphicx}
\usepackage{cite}
\usepackage{hyperref}
\usepackage{ulem}
\advance \textheight by \topskip
\def\*{{\varstar}}

\def\be{\begin{equation}}
\def\ee{\end{equation}}
\def\bea{\begin{eqnarray}}
\def\eea{\end{eqnarray}}

\newcommand{\ie}{\textit{i.e. }}
\newcommand{\eg}{\textit{eg.}}

\begin{document}
	
\title{Electromagnetically and gravitationally stealth fields}


\renewcommand\Authands{ and }
\author[1]{\sf Cristian Quinzacara\thanks{cristian.quinzacara at uss.cl}}

\author[1]{\sf Paola Meza\thanks{paola.meza at uss.cl.}}

\author[2]{\sf Almeira Sampson\thanks{almeira.sampson at uss.cl.}}

\author[2,3]{\sf Mauricio Valenzuela\thanks{mauricio.valenzuela at uss.cl}}

\affil[1]{\textit{Facultad de Ingenier\'ia, Arquitectura y Dise\~no,}\authorcr
	\textit{Universidad San Sebasti\'an, Concepci\'on, Chile.
 }\vspace{1.5ex}}

\affil[2]{\textit{Facultad de Ingenier\'ia, Arquitectura y Dise\~no,}\authorcr
	\textit{Universidad San Sebasti\'an, Valdivia, Chile.
 }\vspace{1.5ex}}

\affil[3]{\textit{Centro de Estudios Cient\'{\i}ficos (CECs), Arturo Prat 514, Valdivia, Chile}\vspace{1.5ex} }
	\date{}
	
	\maketitle

	\begin{abstract}

We construct a generic class of models for complex scalar fields---minimally coupled to gravity and electromagnetism---with the property that their energy-momentum tensor and the electric current vanish for certain massive configurations. These are electromagnetically and gravitationally  {\it stealth fields}. We shall see that the latter configurations can affect, in addition, the strength of the gravity-matter and electromagnetic-matter couplings of other (non-stealth) modes present in the system, which turn out to be equivalent to the re-scaling the electric charge and the Newton constant (with a stealth-mass depending factor).
		
	\end{abstract}

	\newpage
	
	\tableofcontents
	
	\section{Introduction}

It has been shown by many authors \cite{AyonBeato:2001sb,Henneaux:2002wm,Gegenberg:2003jr,AyonBeato:2004ig,AyonBeato:2005tu,Martinez:2005di,Hassaine:2006gz,Faraoni:2010mj,Maeda:2012tu,Ayon-Beato:2013bsa,Hassaine:2013cma,Babichev:2013cya,Babichev:2015rva,Ayon-Beato:2015qfa,Chagoya:2017ojn,Ayon-Beato:2015mxf,Alvarez:2016qky,Smolic:2017bic,Alvarez:2017yzr,Alvarez:2018bkl,Alvarez:2019pth,BenAchour:2018dap, Motohashi:2019ymr,BenAchour:2020wiw} that scalar matter fields may not always curve the spacetime, in spite of their non-trivial degrees of freedom. Similar results were obtained in the context of massive gauge theories in three dimensions \cite{Oliva:2014pia}. This type of fields are now dubbed {\it stealth fields}. These curious objects deserve some attention since, besides of being counter examples of the idea that matter always curves the space, they could also play a role in black-hole physics \cite{AyonBeato:2004ig,Hassaine:2013cma,Babichev:2013cya,Babichev:2015rva,Smolic:2017bic,BenAchour:2018dap,BenAchour:2020wiw}, in cosmology \cite{Maeda:2012tu,Ayon-Beato:2013bsa,Ayon-Beato:2015mxf,Alvarez:2018bkl,Alvarez:2019pth}, and in the description of neutron stars  \cite{Cisterna:2015yla,Cisterna:2016vdx}, among other applications.

The existence of stealth field configurations indicates that the spacetime curvature may not reflect the presence of matter as expected from standard field theories in curved space. Indeed, it is well known that the gravitational Hilbert energy-momentum tensor of the visible matter in galaxies, described by standard field theories, fails to explain the observed spacetime curvature, which has lead to the proposal of {\it dark matter}. Alternatively, one could argue that perhaps the field theory describing the matter at large scales should be modified.

Extending the results of reference \cite{Quinzacara:2018kyn}, we consider a generic action principle for a complex scalar field coupled to gravity and electromagnetism. According to a determined procedure we obtain a new {\it deformed action principle}, which is a uni-parametric extension of the original model containing stealth fields.  The respective equations of motion, which are in general of higher order in derivatives, may contain non-stealth solutions inherited the original theory. For the stealth configurations both, the energy-momentum tensor and the electric current vector vanish. Interestingly, the  energy-momentum tensor of the non-stealth configurations as well as their electric currents, in the deformed theory, are proportional to the ones of the original model with factors that depend on the stealth field mass parameter.

The original and the deformed theories are not equivalent, since the deformation increases the number of degrees of freedom. The deformation method is just a way to build a theory with the mathematical property of possessing stealth solutions.

For example, for the minimally coupled Klein-Gordon complex field, of mass $M$, our procedure yields a 6th order in derivatives action principle,  which contains as solution the stealth configurations of mass $\theta^{-1}$ (being $\theta$ the deformation parameter) and it preserves the solution of the original model of mass $M$. When the Hilbert energy-momentum tensor is obtained, we observe that it vanishes for the stealth fields, and the same happens for the electric current. However, for the mass $M$ configuration the energy-momentum tensor and the electric current are rescaled with respect to the original Klein-Gordon action by a factor $1-(M\theta)^2$.  This is, the effects of massive scalar fields on their gravity and electromagnetic background can be fine-tuned using the stealth field mass $\theta^{-1}$.

This paper is organized as follows. In section \ref{sec:stealthdefinition} we introduce our notation and define what a stealth field is. In section \ref{sec:redefinition} we define the deformation of the action principle and obtain the respective equations of motion, as deformations of the original equations of motion. We prove that the deformed theories contain massive fields with mass inversely proportional to the deformation parameter. In section \ref{sec:examples} we construct some examples and characterize their solutions and in section \ref{Overview} we present some conclusions.

\section{Stealth fields} \label{sec:stealthdefinition}

In this section we shall follow closely reference \cite{Quinzacara:2018kyn}. This is, in   $D$-dimensional spacetime dimensions, with dynamical metric $g_{\mu\nu}$ with signature $(-,+,+,...)$, consider the action principle,
\be\label{Sgphi}
S[g,A,\phi,\widebar\phi] = S_{G}[g]+S_{A}[g,A] + S_M[g,A,\phi,\widebar\phi],
\ee
where $S_{G}[g]$ is the  purely gravitational sector,  $S_{A}[g,A]$ and $S_M[g,A,\phi,\widebar\phi]$ are respectively the terms of the $U(1)$ gauge field and the complex scalar matter field minimally coupled to $U(1)$, both in the geometric background provided by the metric $g$. Here $\widebar\phi=\phi^*$ is the complex conjugated of $\phi$. We do not need to specify the details of each sector, but
under gauge transformation, with parameter $\alpha(x)$, the fields must transform as,
\be\label{U1}
\begin{array}{c}
\gamma=\exp(i\alpha(x)) \;\in\:U(1)\\[5pt]
\phi \rightarrow \gamma\phi,\quad \widebar \phi \rightarrow \widebar \phi \gamma^*,\quad A_\mu\rightarrow A_\mu-i\partial_\mu \alpha\,.
\end{array}
\ee
and hence, the spatial derivatives must appear covariantly as, \be\label{cd01}
D_\mu \phi:=(\nabla_\mu+i A_\mu)\phi, \qquad \widebar D_\mu \bar \phi:=(\nabla_\mu-i A_\mu)\bar\phi,
\ee
 where $\nabla_\mu$ is the Levi-Civita derivative.

The variation of this action with respect to the (inverse) metric tensor $g^{\mu\nu}$ yields,
\be\label{var-g}
\frac{\delta S[g,A,\phi,\widebar\phi]}{\delta g^{\mu\nu}}= \sqrt{-g}\Bigl(H_{\mu\nu}[g]-\Theta_{\mu\nu}[g,A] -\Xi_{\mu\nu}[g,A,\phi,\widebar\phi]\Bigr),
\ee
where 
\begin{eqnarray} 
H_{\mu\nu}[g] &:=& \frac{1}{\sqrt{-g}}\frac{\delta S_{G}[g]}{\delta g^{\mu\nu}} ,\label{H}\\[5pt]
 \Theta_{\mu\nu}[g,A] &:=& -\frac{1}{\sqrt{-g}}\frac{\delta S_A[g,A]}{\delta g^{\mu\nu}}, \label{Theta} \\[5pt] 
\Xi_{\mu\nu}[g,A,\phi,\widebar\phi]&:=& -\frac{1}{\sqrt{-g}}\frac{\delta S_M[g,A,\phi,\widebar\phi]}{\delta g^{\mu\nu}} \,, \label{Xi}
\end{eqnarray}
are respectively the (generalized) Einstein tensor, the electromagnetic and the scalar field  energy-momentum tensors. 
	
From the variation of the action with respect to the scalar fields $\phi$ and $\widebar\phi$ we define, 
\be \label{Upsilon}
\Upsilon[g,A,\phi,\widebar\phi]:= \frac{\delta S_M[g,A,\phi,\widebar\phi]}{\delta \phi} 
\,,
\ee
\be \label{Upsilon*}
\widebar\Upsilon[g,A,\phi,\widebar\phi]:= \frac{\delta S_M[g,A,\phi,\widebar\phi]}{\delta \widebar\phi} 
\,.
\ee
Note that  \eqref{Upsilon}  and \eqref{Upsilon*} transforms under the $U(1)$ transformations \eqref{U1} as,
\be
\begin{array}{c}
\widebar\Upsilon[g,A,\phi,\widebar\phi] \rightarrow \gamma\widebar\Upsilon[g,A,\phi,\widebar\phi] ,\\[5pt]
\Upsilon[g,A,\phi,\widebar\phi]\rightarrow \Upsilon[g,A,\phi,\widebar\phi] \gamma^*\,.
\end{array}
\ee

The variation of the action with respect to the gauge field $A_\mu$ yields,
\be\label{var-A}
\frac{\delta S[g,A,\phi,\widebar\phi]}{\delta A_\mu}= \sqrt{-g}\Bigl(-E^\mu[g,A]+J^\mu[g,A,\phi,\widebar\phi]\Bigr),
\ee
where we have defined 
\be\label{E}
E^\mu[g,A]:=-\frac{1}{\sqrt{-g}}\frac{\delta S_A[g,A]}{\delta A_\mu}
\ee
and the $U(1)$ gauge current 
\be\label{J}
J^\mu[g,A,\phi,\widebar\phi]:=\frac{1}{\sqrt{-g}}\frac{\delta S_M[g,A,\phi,\widebar\phi]}{\delta A_\mu}\,.
\ee

Using these definitions, the field equations of the theory \eqref{Sgphi} read,
\be \label{eomgphi}
H_{\mu\nu} [g]-\Theta_{\mu\nu}[g,A]-\Xi_{\mu\nu}[g,A,\phi,\widebar\phi] =0\,,
\ee
\be \label{eomphi}
\Upsilon[g,A,\phi,\widebar\phi]=0\, ,\qquad\widebar\Upsilon[g,A,\phi,\widebar\phi]=0\,,
\ee
\be \label{eomA}
E^\mu[g,A]-J^\mu[g,A,\phi,\widebar\phi]=0\,.
\ee

A stealth complex scalar field is a non trivial solution of the field equations \eqref{eomphi} which in addition has a vanishing energy-momentum tensor, 
\be
\Xi_{\mu\nu}[g,A,\phi,\widebar\phi]=0\,.
\ee
Hence, from \eqref{eomgphi}, the generalized Einstein tensor is sourced only by the $U(1)$ gauge energy-momentum tensor,
\be 
H_{\mu\nu}[g]=\Theta_{\mu\nu}[g,A]\,,
\ee 
so that in presence of the stealth fields $\phi$ and $\widebar{\phi}$ the metric tensor must satisfy identical equations of motion than in the vacuum $\phi=\widebar\phi=0$ of the scalar fields. 
	
\section{$\theta$-deformation of complex scalar field theory}\label{sec:redefinition}

 Consider a theory given by the action principle $S[g,A,\phi,\widebar\phi]$. We produce a new model by means of the deformation map:
 \be \label{deformation}
S[g,A,\phi,\widebar\phi]  \quad \rightarrow \quad S^\theta[g,A,\phi,\widebar\phi]:= S[g,A,\phi^\theta, \widebar\phi^\theta]\,,
\ee
where 
\be \label{phitheta}
\phi^\theta[g,A,\phi] = (1-\theta^2 D^2)\phi\, ,\qquad \widebar\phi^{\theta}[g,A,\widebar\phi] = (1-\theta^2 \widebar D^2)\widebar\phi\,,
\ee
 and $\theta$ is a real-valued parameter with dimensions of $[length]^{-1}$. 
 
Thus, from the ``original" action principle $S[g,A,\phi,\widebar\phi]$, we obtain its {\it deformation}  $S^\theta[g,A,\phi,\widebar\phi]$, by means of the formal replacement,
\be\label{phiphitheta}
\phi \: \rightarrow \: \phi^\theta\,, \qquad \widebar\phi\: \rightarrow \: \widebar\phi^\theta\,,
\ee 
which only affects the matter sector. Hence we define, 
\be \label{defSMAction}
S^\theta_M[g,A,\phi,\widebar\phi]:=S_M\bigl[g,A,\phi^\theta[g,A,\phi],\widebar\phi^\theta[g,A,\widebar\phi]\bigr]\,.
\ee

Observe that the deformed action principle takes the same value
\be
S_M\bigl[g,A,\phi^\theta[g,A,\phi],\widebar\phi^\theta[g,A,\widebar\phi]\bigr]\Bigg|_{\phi^\theta=0}=S_M\bigl[g,A,\phi,\widebar\phi\bigr]\Bigg|_{\phi=0},
\ee 
on the solution of the minimally coupled Klein-Gordon equation,
 \be \label{KG}
\phi^\theta[g,A,\phi_m] = -\theta^2 (D^2-\theta^{-2})\phi_m=0\,,
 \ee
than the original theory valued in the trivial solutions $\phi=0$. Thus, assuming that the original action takes an extreme value in the trivial vacuum $\phi=0$, it must be also true that the deformed action takes extreme value in the massive configuration $\phi=\phi_m$.

\subsection{Field equations}

The equations of motion of deformed theory \eqref{deformation}, 
\be \label{deltaSgtheta}
\begin{array}{l}
\dfrac{\delta S^\theta[g,A,\phi,\widebar\phi]}{\delta g^{\mu\nu}}= 0,\quad \dfrac{\delta S^\theta[g,A,\phi,\widebar\phi]}{\delta A_\mu} = 0,\\[5pt]
\dfrac{\delta S^\theta[g,A,\phi,\widebar\phi]}{\delta \phi} = 0,\quad \dfrac{\delta S^\theta[g,A,\phi,\widebar\phi]}{\delta \widebar\phi} = 0,
\end{array}
\ee
are given by,
\be \label{defeomgphi}
\begin{array}{l}
H_{\mu\nu} [g]- \Theta_{\mu\nu}[g,A]-\widetilde\Xi_{\mu\nu}[g,A,\phi,\widebar\phi] =0,\\[5pt]
 E^{\mu}[g,A]-\widetilde J^{\mu}[g,A,\phi,\widebar\phi]=0,
\end{array}
\ee
\be \label{defeomgphi2}
\widetilde\Upsilon[g,A,\phi,\widebar\phi]=0,\qquad,\widetilde{\widebar{\Upsilon}}[g,A,\phi,\widebar\phi]=0,
\ee
using the definitions \eqref{H}, \eqref{Theta}, \eqref{E}, and 
\begin{equation} \label{EMtensortheta}
\widetilde\Xi_{\mu\nu}[g,A,\phi,\widebar\phi]:=-\frac{1}{\sqrt{-g}}\frac{\delta S^\theta_M[g,A,\phi,\widebar\phi]}{\delta g^{\mu\nu}} \,,
\end{equation}
\be \label{currenttheta}
\widetilde J^\mu[g,A,\phi,\widebar\phi]:=\frac{1}{\sqrt{-g}}\frac{\delta S^\theta_M[g,A,\phi,\widebar\phi]}{\delta A_\mu} \,,
\ee
\be \label{Upsilontheta}
\begin{array}{l}
\widetilde{\Upsilon}_{\mu\nu}[g,A,\phi,\widebar\phi]{:=}\dfrac{\delta S^\theta_M[g,A,\phi,\widebar\phi]}{\delta \phi} \,,\\[5pt]\widetilde{\widebar\Upsilon}_{\mu\nu}[g,A,\phi,\widebar\phi]:=\dfrac{\delta S^\theta_M[g,A,\phi,\widebar\phi]}{\delta \widebar\phi}.
\end{array}
\ee
Note that in \eqref{defeomgphi} $H_{\mu\nu} [g]$, $\Theta_{\mu\nu}[g,A]$ and $E^\mu[g,A]$ remain undeformed with respect to  \eqref{H}.
 The tildes on $\widetilde{\Xi}$, $\widetilde{J}$ and $\widetilde{\Upsilon}$ and $\widetilde{\widebar\Upsilon}$ indicate that these magnitudes are analogous to the original variables \eqref{Xi},  \eqref{Upsilon}, \eqref{Upsilon*} and \eqref{J}, in the deformed theory \eqref{defSMAction}.

The results of the variation of the deformed action principle \eqref{deformation} with respect to the matter fields and the spacetime metric can be related to the results of the undeformed action principle \eqref{Sgphi} using the chain rule of functionals,
\be\label{FG}
\frac{\delta F[G[f]]}{\delta f(y)}=\int d^{D}z \frac{\delta F[G[f]]}{\delta G[f](z)} \frac{\delta G[f](z)}{\delta f(y)}\, ,
\ee
for two functionals $F[f]$ and $G[f]$ of the function $f$ and their composition $F[G[f]]$. We declare the dependence of a functional on the point $x$ by attaching the symbol $(x)$ to it, \eg  \,$F[f](x)$.  Applying this in the computation of the equations of motion \eqref{Upsilontheta}, in the point $y$, we obtain from the deformed action, 
\be
\frac{\delta S_{M}^{\theta}[g,A,\phi,\widebar\phi]}{\delta\phi(y)}=\int d^{D}z \frac{\delta S^\theta_{M}[g,A,\phi,\widebar\phi]}{\delta\phi^\theta(z)}\frac{\delta\phi^\theta(z)}{\delta\phi(y)}\,,
\ee	
in account of \eqref{defSMAction}. The latter expression is equivalent to,
\be\label{deltaSMphi}
\frac{\delta S_{M}^{\theta}[g,A,\phi,\widebar\phi]}{\delta\phi(y)}=\int d^{D}z\,  \Upsilon^\theta[g,A,\phi,\widebar\phi](z)\frac{\delta\phi^\theta(z)}{\delta\phi(y)}\,,
\ee
where we define
\begin{equation}\label{Uphi1}
 \Upsilon^\theta[g,A,\phi,\widebar\phi]:=\Upsilon[g,A,\phi^\theta,\widebar\phi^\theta]\,,
\end{equation}
and  
\be\label{Uphi2}
\frac{\delta S_{M}[g,A,\phi^\theta,\widebar\phi^\theta]}{\delta\phi^\theta(z)}=\frac{\delta S_{M}[g,A,\phi,\widebar\phi]}{\delta\phi(z)}\Biggr|_{\substack{\phi \rightarrow \phi^{\theta}[g,A,\phi]\\\widebar\phi \rightarrow \widebar\phi^{\theta}[g,A,\widebar\phi]}} \,, 
\ee
is equivalent to the original operator \eqref{Upsilon} valued in  $\phi^{\theta}[g,A,\phi]$ and $\widebar\phi^{\theta}[g,A,\widebar\phi]$.
Note that in \eqref{deltaSMphi},
\be
\frac{\delta\phi^\theta(z)}{\delta\phi(y)}=(1-\theta^{2}D^2)\delta^{D}(z-y)\,.
\ee
Integrating by parts and with the definitions \eqref{Upsilontheta} and \eqref{Uphi1}
we obtain the equation of motion for the matter field in the deformed theory \eqref{defeomgphi},
\be\label{Upsilontheta2}
\widetilde\Upsilon[g,A,\phi,\widebar\phi]:=(1-\theta^{2}\widebar{D}^2)\Upsilon^{\theta}[g,A,\phi,\widebar\phi]=0\,,
\ee
\be\label{barUpsilontheta2}
\widetilde{\widebar{\Upsilon}}[g,A,\phi,\widebar\phi]:=(1-\theta^{2}{D}^2)\widebar\Upsilon^{\theta}[g,A,\phi,\widebar\phi]=0\,,
\ee
where 
\be\label{barUphi1}
 \widebar{\Upsilon}^\theta[g,A,\phi,\widebar\phi] := \widebar{\Upsilon}[g,A,\phi^\theta,\widebar\phi^\theta]\,.
\ee

The variation of the action with respect to $g^{\mu\nu}$ should be carried out taking similar considerations. For this purpose, let us introduce the Lagrangian density $\mathcal{L}_M[g,A,\phi,\widebar\phi]$, such that,
\be\label{SL}
S_{M}[g,A,\phi,\widebar\phi]=\int d^{D}x\sqrt{-g}\mathcal{L}_M[g,A,\phi,\widebar\phi],
\ee
which with the substitution \eqref{defSMAction} yields,
\be\label{defMS}
S_{M}^{\theta}[g,A,\phi,\widebar\phi]=\int d^{D} x\sqrt{-g}\mathcal{L}_M[g,A,\phi^\theta,\widebar\phi^\theta]\, .
\ee
From the chain rule  \eqref{FG}, considering that the deformed action functional depends on $g^{\mu\nu}$ explicitly and also implicitly in the deformed fields $\phi^\theta$ and $\widebar\phi^\theta$, the variation of \eqref{defSMAction} with respect to the metric is equivalent to,
\bea\label{VardefMS1}
\frac{\delta S_{M}[g,A,\phi^\theta,\widebar\phi^\theta]}{\delta g^{\mu\nu}(y)} &=& \frac{\delta S_{M}[g,A,\phi,\widebar \phi]}{\delta g^{\mu\nu}(y)} \Bigg\vert_{\substack{\phi \rightarrow \phi^{\theta}\\\widebar\phi \rightarrow \widebar\phi^{\theta}}} \\ &&+ \int d^{D}z \frac{\delta S_{M}[g,A,\phi^\theta,\widebar\phi^\theta]}{\delta \phi^\theta(z)}  \frac{\delta \phi^\theta(z)}{\delta g^{\mu\nu}(y)}\nonumber + \int d^{D}z \frac{\delta S_{M}[g,A,\phi^\theta,\widebar\phi^\theta]}{\delta \widebar\phi^\theta(z)}  \frac{\delta \widebar\phi^\theta(z)}{\delta g^{\mu\nu}(y)}\,.\nonumber
\eea
Considering the definitions \eqref{Uphi1}, \eqref{barUphi1} and \eqref{defMS}, \eqref{VardefMS1} can be written also as,
\bea\label{VardefMS2}
\frac{\delta S^\theta_{M}[g,A,\phi,\widebar\phi]}{\delta g^{\mu\nu}(y)}& =& - \sqrt{-g} \, \Xi^\theta_{\mu\nu}  (y)+ \int d^{D}z\, \Upsilon^\theta(z) \frac{\delta \phi^\theta(z)}{\delta g^{\mu\nu}(y)}+ \int d^{D}z\, \widebar{\Upsilon}^\theta(z) \frac{\delta \widebar{\phi}^\theta(z)}{\delta g^{\mu\nu}(y)} \,,\nonumber
\eea
where 
\be\nonumber
\Xi^\theta_{\mu\nu}:=\Xi_{\mu\nu} [g,A,\phi^\theta,\widebar\phi^\theta] \,.
\ee

From \eqref{VardefMS2}  the energy-momentum tensor of the scalar field in deformed theory \eqref{EMtensortheta} is given by,
\bea\label{EMtensortheta2}
\widetilde\Xi_{\mu\nu}[g,\phi] (y)&=&  \Xi^\theta_{\mu\nu} (y)-\frac{1}{\sqrt{-g}} \int d^{D}z \Upsilon^\theta(z) \frac{\delta \phi^\theta(z)}{\delta g^{\mu\nu}(y)}-\frac{1}{\sqrt{-g}} \int d^{D}z \widebar\Upsilon^\theta(z) \frac{\delta \widebar\phi^\theta(z)}{\delta g^{\mu\nu}(y)}
\,.\nonumber
\eea
where
\bea
\frac{\delta\phi^\theta(z)}{\delta g^{\mu\nu}(y)}&=& 
 -\frac{1}{2}\Big(g_{\mu\nu} g^{\sigma\rho} (D_\sigma\phi) (\partial_\rho \delta^D(z-y)\Big) +\frac{1}{\sqrt{-g}} \Big(\partial_{(\mu}+iA_{(\mu}\Big)\Big(\sqrt{-g}\delta^D(z-y)D_{\nu)}\phi\Big)\,,\nonumber
\eea
and
\be
\frac{\delta\widebar\phi^\theta(z)}{\delta g^{\mu\nu}(y)} = \left(\frac{\delta\phi^\theta(z)}{\delta g^{\mu\nu}(y)}\right)^*\,,
\ee
and we have used the conventional notation for index symmetrization (with factor $1/2$).

Substituting in \eqref{EMtensortheta2} followed by an integration by parts, and considering the boundary conditions

\be
\Upsilon^\theta[g,A,\phi,\widebar \phi ]\Big\vert_{\partial M}=\widebar\Upsilon^\theta[g,A,\phi,\widebar \phi ]\Big\vert_{\partial M}=0\,,
\ee
we obtain the energy-momentum tensor of deformed theory,
\bea\label{temtheta} 
\widetilde\Xi_{\mu\nu}&=&\Xi^{\theta}_{\mu\nu}+\frac{\theta^{2}}{2}\frac{1}{\sqrt{-g}}g_{\mu\nu}\left((D^2\phi) \Upsilon^\theta+(\widebar D^2\widebar\phi) \widebar\Upsilon^\theta\right)\label{EMtensortheta3}\\
&&-\frac{\theta^{2}}{2}(\delta_\mu^\rho\delta_\nu^\sigma+\delta_\mu^\sigma\delta_\nu^\rho-g_{\mu\nu}g^{\sigma\rho})\left( (D_\rho\phi )\widebar D_\sigma\left(\frac{\Upsilon^\theta}{\sqrt{-g}}\right)
+(\widebar D_\rho \widebar \phi ) D_\sigma\left(\frac{\widebar\Upsilon^\theta}{\sqrt{-g}}\right)\right)\,.\nonumber
\eea
Following an analogously procedure, the computation of the electric current \eqref{currenttheta} of the deformed theory \eqref{defSMAction} yields,
\bea
\widetilde{J}^{\mu}[g,A,\phi,\widebar\phi] & = & J^{\mu}[g,A,\phi^\theta,\widebar{\phi}^\theta]   -i\frac{\theta^2}{\sqrt{-g}} \Big( \Upsilon^\theta D^\mu \phi - \widebar{\Upsilon}^\theta \widebar{D}^\mu \widebar\phi- \phi  \widebar D^\mu\Upsilon^\theta  + \widebar\phi{D}^\mu\widebar{\Upsilon}^\theta \Big)\label{Jtilde}.
\eea
The  first term on the r.h.s. of \eqref{Jtilde} appears from the variation of the part of the action which depends explicitly on the gauge field $A_\mu$ and it is equivalent to electric current functional of the original theory \eqref{J} valued in  $\phi^\theta$ \eqref{phitheta} instead of  $\phi$. The remaining terms appear from the implicit dependence of $\phi^\theta$ in the gauge field $A_\mu$ through \eqref{phitheta}.

\subsection{Stealth theorem}\label{Stheo}

The deformation method insures that all obtained models contain stealth configurations. To see this, consider the solutions $\phi=\phi_m$ and $\widebar\phi=\widebar\phi_m$ of mass $m=\theta^{-1}$ the Klein-Gordon equations \eqref{KG}, such that 
\be\label{KG2}
\phi^{\theta}[g,A,\phi_m]=\widebar\phi^{\theta}[g,A,\widebar \phi_m]=0.
\ee
They are solutions of the deformed theory \eqref{defeomgphi2},
\bea \label{eom1}
\widetilde\Upsilon[g,A,\phi_m,\widebar\phi_m]&=&(1-\theta^{2}D^2)\Upsilon^{\theta}[g,A,\phi_m,\widebar\phi_m]\\&=&(1-\theta^{2}D^2)\Upsilon[g,A,0,0]=0\,,\nonumber
\eea
\bea \label{eom1c}
\widetilde{\widebar\Upsilon}[g,A,\phi_m,\widebar\phi_m]&=&(1-\theta^{2}\widebar D^2)\widebar\Upsilon^{\theta}[g,A,\phi_m,\widebar\phi_m]\\&=&(1-\theta^{2}\widebar D^2) \widebar \Upsilon[g,A,0,0]=0\,,\nonumber
\eea
 under the assumption that the field equations of the original theory \eqref{eomphi} admit the trivial solutions $\phi=\widebar\phi=0$, namely 
\bea\label{up0}
\Upsilon^{\theta}[g,A,\phi_m,\widebar\phi_m]=\Upsilon[g,A,0,0]=0,\qquad
\widebar\Upsilon^{\theta}[g,A,\phi_m,\widebar\phi_m]=\widebar\Upsilon[g,A,0,0]=0\,.
\eea

Now, let us evaluate the energy-momentum tensor \eqref{temtheta} in the same set of  solutions,
\bea 
\widetilde\Xi_{\mu\nu}[g,A,\phi_m,\widebar\phi_m]&=&\Xi^\theta_{\mu\nu}[g,A,\phi_m,\widebar\phi_m]\\&+&\frac{\theta^{2}}{2}\frac{g_{\mu\nu}}{\sqrt{-g}}\Big\{ (D^2\phi_m) \Upsilon^\theta[g,A,\phi_m,\widebar\phi_m]+(\widebar D^2\widebar\phi_m) \widebar\Upsilon^\theta[g,A,\phi_m,\widebar\phi_m]\Big\} \nonumber\\
&&-\frac{\theta^{2}}{2}(\delta_\mu^\rho\delta_\nu^\sigma+\delta_\mu^\sigma\delta_\nu^\rho-g_{\mu\nu}g^{\sigma\rho})\Big\{ (D_\rho\phi_m )\widebar D_\sigma\left(\frac{\Upsilon^\theta[g,A,\phi_m,\widebar\phi_m ]}{\sqrt{-g}}\right)\nonumber \\&&
\hspace{5cm}+(\widebar D_\rho \widebar \phi_m ) D_\sigma\left(\frac{\widebar\Upsilon^\theta[g,A,\phi_m,\widebar\phi_m ]}{\sqrt{-g}}\right)\Big\}\,. \nonumber
\eea
This tensor vanishes from \eqref{up0} and since 
\bea
\Xi^\theta_{\mu\nu}[g,A,\phi_m,\widebar\phi_m]&=&\Xi_{\mu\nu}[g,A,\phi^\theta[g,A,\phi_m],\widebar\phi^\theta[g,A,\widebar\phi_m]]\nonumber\\&=&\Xi_{\mu\nu}[g,A,0,0]\,,
\eea
equals the scalar energy momentum tensor of the original theory in the vacuum of the scalar field, it  must also vanish. It follows that the configurations $\phi=\phi_m$ are stealth.

Similarly, the electric current \eqref{Jtilde} vanishes as consequence of \eqref{up0},
\bea
\widetilde J^{\mu}[g,A,\phi_m,\widebar\phi_m]&=&  J^{\mu}[g,A,\phi^\theta[g,A,\phi_m],\widebar\phi^\theta[g,A,\widebar\phi_m]\nonumber\\&=&J^{\mu}[g,A,0,0]=0\,,
\eea
since the current of the original theory valued at the trivial vacuum of the scalar field must vanish. 

Finally, in the deformed theory \eqref{deformation} the system of equations \eqref{defeomgphi}-\eqref{defeomgphi2}, are reduced to the Einstein equations in the presence of a electromagnetic fields and the Maxwell equations in a curved background without matter source,
\be \label{defeomgphi3}
H_{\mu\nu} [g]- \Theta_{\mu\nu}[g,A] =0, \qquad E^{\mu}[g,A]=0\,,
\ee
{\it i.e.} the same than the equations of motion of the original theory \eqref{Sgphi} without matter field, $S_M[g,A,\phi,\widebar\phi]=0$. Therefore the mass $\theta^{-1}$ solution $\phi_m$, is stealth in two ways: it does not source space-time curvature nor electromagnetic field.

\section{Complex-Klein Gordon case}\label{sec:examples}

The deformation map \eqref{deformation} apply to any complex scalar theory, with the only requirement that the theory $S[g,A,\phi,\widebar\phi]$ has the trivial vacuum $\phi=0$ as solution. Then there is a unique deformation $S^\theta[g,A,\phi,\widebar\phi]$ that contains a stealth field configuration.

To put these ideas into practice, let us consider the minimally coupled Klein-Gordon action of a complex field of mass $M$,
\begin{equation}\label{EX:action}
S_M[g,A,\phi,\widebar\phi]=-\int d^Dx\, \sqrt{-g}\,\left(g^{\mu\nu}D_\mu\phi \widebar D_\nu\widebar \phi+M^2\phi\widebar \phi\right)\,,
\end{equation}
with the covariant derivatives defined as in \eqref{cd01}. The field equations are given by,
\begin{equation}\label{EX:fe}
\begin{array}{l}
\Upsilon[g,A,\phi,\widebar\phi]=\sqrt{-g}\,\left(\widebar D^2-M^2\right)\widebar\phi=0\,,\\[5pt] 
\widebar \Upsilon[g,A,\phi,\widebar\phi]=\sqrt{-g}\,\left( D^2-M^2\right)\phi=0\,.
\end{array}
\end{equation}
The electric current \eqref{J} is given by,
\begin{equation}\label{EX:ec}
J^\mu=i\left(\widebar{\phi}D^\mu\phi-\phi\widebar{D}^\mu\widebar{\phi}\right)\,,
\end{equation}
and the energy-momentum tensor $\Xi_{\mu\nu}$ of the scalar field is
\begin{equation}\label{EX:emt}
\Xi_{\mu\nu}=\frac{1}{2}\Bigl\{g_{\mu\nu}\left(D^{\rho}\phi \widebar D_{\rho}\widebar\phi+M^2\phi\widebar\phi \right)-\left(D_\mu\phi\widebar D_\nu\widebar\phi+  D_\nu\phi\widebar D_\mu\widebar\phi\right)\Bigr\}\,.
\end{equation}

When applied to \eqref{EX:action} the deformation map \eqref{defSMAction} yields,
\bea
S_M^\theta[g,A,\phi,\bar\phi]&=&-\int d^D\sqrt{-g}\left(\small g^{\mu\nu}D_\mu\phi^\theta[g,A,\phi] \widebar D_\nu\widebar \phi^\theta[g,A,\widebar\phi]+M^2\phi^\theta[g,A,\phi]\widebar \phi^\theta[g,A,\widebar\phi]\right)\,,\nonumber
\eea
which should be expanded using the definitions  \eqref{phitheta} of $\phi^\theta$ and $\widebar\phi^\theta$. We obtain the field equation, 
\bea\label{Ex:eom1}
\widetilde{\bar\Upsilon}[g,A,\phi,\widebar\phi]&=&\sqrt{-g}\left(1-\theta^2 D^2\right)\left( D^2-M^2\right)\left(1-\theta^2 D^2\right)\phi=0\,,\\
\label{EX:fephi}
&=&\theta^4\sqrt{-g}\left(D^2-\theta^{-2}\right)^2\left( D^2-M^2\right)\phi=0\,,
\eea
which takes the form \eqref{Upsilontheta2}. 

Analogously, the field equations of the conjugated field can be written as,
\begin{equation}\label{EX:febarphi}
\widetilde{\Upsilon}[g,A,\phi,\widebar\phi]=\theta^4\sqrt{-g}\left(\bar D^2-\theta^{-2}\right)^2\left(\bar D^2-M^2\right)\bar \phi=0\,.
\end{equation}

The electric current, as defined in \eqref{Jtilde}, is given by,
\bea\label{EX:ecdef}
\widetilde J^\mu[g,A,\phi,\bar\phi]&=&i\left(\widebar\phi^\theta D^\mu\phi^\theta-\phi^\theta \widebar D^\mu \widebar\phi^\theta\right)\\&&\hspace*{-1cm}-{i \theta^2}\biggl\{ D^\mu\phi\, \left(\widebar D^2-M^2\right)\widebar\phi^\theta - \left(D^2-M^2\right)\phi^\theta\widebar D^\mu\widebar\phi -\phi\,\widebar D^\mu\left(\widebar D^2-M^2\right)\widebar\phi^\theta+\widebar\phi D^\mu\left(D^2-M^2\right)\phi^\theta\biggr\}\,.\nonumber
\eea

The energy-momentum tensor  \eqref{EMtensortheta} of deformed matter yields in this case,
\bea\label{EX:temdef}
\widetilde\Xi_{\mu\nu}[g,A,\phi,\bar\phi]&=&\frac{1}{2}\Bigl\{g_{\mu\nu}\left(D^{\rho}\phi^\theta \bar D_{\rho}\bar\phi^\theta+M^2\phi^\theta\bar\phi^\theta \right)\\&&-\left(D_\mu\phi^\theta\bar D_\nu\bar\phi^\theta+  D_\nu\phi^\theta\bar D_\mu\bar\phi^\theta\right)\Bigr\}+\frac{\theta^2}{2}g_{\mu\nu}\Bigl\{D^2\phi\left(\bar D^2-M^2\right)\bar\phi^\theta+\left(D^2-M^2\right)\phi^\theta\bar D^2\bar\phi\Bigr\}\,.\nonumber
\eea

The stealth field satisfies, 
\begin{equation}
\phi^\theta=0=\left(1-\theta^{2}D^2\right)\phi_m=-\frac{1}{m^2}\left(D^2-m^2\right)\phi_m\,,
\end{equation}
with mass $m=\theta^{-1}$. It is straightforward to show that the field equations \eqref{EX:fephi}-\eqref{EX:febarphi} are satisfied and that both the electric current \eqref{EX:ecdef} and energy-momentum tensor \eqref{EX:temdef} vanish for $\phi=\phi_m$. Hence, though non-trivial, $\phi_m$ does not curve the spacetime and it does not source electromagnetic fields.

\subsection*{Stealth fields as modifiers of gravitational and electromagnetic couplings of regular configurations}

From the factorization of the operators in \eqref{EX:fephi}-\eqref{EX:febarphi}, the field equations admit also solutions of mass $M$, 
\begin{equation}\label{EX:KGM}
\left(D^2-M^2\right)\phi_M=0\,,
\end{equation}
which are inherited from the original theory \eqref{EX:fe}. 
Upon these solutions, the the functional $\phi^\theta$ \eqref{phitheta} takes the value, 
\begin{equation}
\phi^\theta[g,A,\phi_M,\bar\phi_M]=-\frac{1}{m^2}\left(D^2-m^2\right)\phi_M=\lambda\phi_M\,,\quad \lambda:=\left(1-(\theta M)^2\right)\,,
\end{equation}
while the electric current \eqref{EX:ecdef} reduces to,
\bea
\widetilde J^\mu[g,A,\phi_M,\bar\phi_M]&=&i\lambda^2\left(D^\mu\phi_M\, \bar\phi_M-\phi_M \bar D^\mu \bar\phi_M\right)\nonumber\\&&=\lambda^2 J^\mu[g,A,\phi_M,\bar\phi_M]\,.\label{EX:ecdefor} 
\eea
Analogously for the conjugated field $\widebar\phi$.

As we can see, in the deformed theory the electric current is re-scaled with respect to its counterpart in the original theory \eqref{EX:ec}. The same behavior is observed in the energy-momentum tensor of deformed theory. Replacing $\phi_M$ ($\widebar\phi_M$) in \eqref{EX:temdef}, we obtain
\begin{equation}\label{EX:temdefor}
\widetilde\Xi_{\mu\nu}[g,A,\phi_M,\bar\phi_M]=\lambda^2 \Xi_{\mu\nu}[g,A,\phi_M,\bar\phi_M]\,.
\end{equation}

We conclude that the general solution of the deformed theory consists on the superposition of the solutions of the original theory and the stealth field. The solutions of the original model have similar electromagnetic and gravitational effect in the deformed theory, though with coupling constant re-scaled by a factor $\lambda^2$. Hence, though stealth fields do not have direct gravitational and electromagnetic effects, they can modify the strength of the effects of regular massive configurations.

\section{Overview and remarks}\label{Overview}

In this paper we observed that there exist a wide class of complex scalar field theories, coupled to gravity and to electromagnetism, that admit stealth configurations. These theories are constructed by means of a ``deformation method'', which can be regarded as a $\theta$-parametric extension of some ``original theory''. We show that the details of the original theory are not important for the existence of stealth configurations in the deformed theories, except that it must admit the trivial vacuum solution.

As a novel aspect, the stealth field in the model presented here, besides not curving the spacetime, does not source electromagnetic fields either. The stealth fields can, however, modify the strength of the effects of regular (non-stealth) scalar fields with respect to the undeformed model. 
In the particular example of the complex Klein-Gordon field of mass $M$, the deformed theory is  of sixth order. There the energy-momentum tensor and the electric current of the original scalar fields are re-scaled by a constant factor $(1-M^2/m^2)^2$, which has the same effect than to re-normalize the Newton constant and the electric charge in the undeformed theory. Hence, the mass of the stealth field $m=\theta^{-1}$, \ie the inverse of the deformation parameter $\theta$, may  be used to diminish/amplify the effects of regular non-stealth fields. 

The deformation map, which involves higher derivatives, results in a different theory. On the one hand, from the presence of time derivatives in its definition \eqref{phitheta}, the deformation map increases the number of propagating degrees of freedom (which can be also seen in the Hamiltonian framework). On the other hand, the deformation map cannot be treated as a regular field redefinition. In the quantum level, the path integral of the deformed theory would not include the Jacobian function, which would be necessary to make the path integral field-parametrization invariant (see \cite{Criado:2018sdb}). Moreover, if introduced, the Jacobian would involve the operator $1-\theta^2 D^2$, which is not invertible (upon the stealth fields).

Concerning ghosts configurations in higher derivative theories, as such, the theories produced by means of the deformation method may also share this problem. Indeed, the models obtained are of the Pais-Uhlenbeck \cite{Pais:1950za}  form, hence they will likely possess Ostrogradsky instabilities. A way to scape to this situation is found in {\it degenerate higher-order scalar-tensor theories} (DHOST) which consists in the elimination of those degrees of freedom by adjustment of the constraint system of the models (see \cite{Langlois:2018dxi} for an interesting review).
We believe that, in spite the potential presence of ghost configurations, the generality of the method here proposed, as well as the non-standard feature of generating electromagnetically stealth fields, justify their considerations while further improvements \textit{\'a la} DHOST, can be seek in the future.

\subsection*{Acknowledgements}

We thank Eloy Ay\'on-Beato, Adolfo Cisterna, and Julio Oliva for their valuable comments. This work was partially funded by grant FONDECYT 1220862.

\end{document}